\documentclass[epj]{svjour}
\usepackage{graphicx}
\usepackage{amssymb}
\usepackage{dcolumn}
\usepackage{amsmath}
\usepackage{bm}
\usepackage{textcomp}
\usepackage{epstopdf}
\usepackage{color,url}
\usepackage{ulem}
\usepackage[toc,page,title,titletoc,header]{appendix}
\usepackage[colorlinks,
            linkcolor=blue,
            anchorcolor=blue,
            citecolor=blue,
            urlcolor=blue
            ]{hyperref}
\usepackage{txfonts}
\usepackage{caption}
\usepackage{lineno}
\usepackage{multirow}
\usepackage{ulem}

\begin{document}
\title{Critical fluctuation patterns and anisotropic correlations in a  non-uniform temperature background}

\author{Lijia Jiang\inst{1,3,4,5} \thanks{{e-mail:} \href{mailto:lijiajiang@nwu.edu.cn}{lijiajiang@nwu.edu.cn} }
\and Tao Yang\inst{2,3,4,5} \and Jun-Hui Zheng\inst{2,3,4,5}
}

\institute{Physics department, Northwest University, Xi'an, 710127, China
\and Institute of Modern Physics, Northwest University, Xi'an, 710127, China
\and Shaanxi Key Laboratory for Theoretical Physics Frontiers, Xi'an, 710127, China
\and Peng Huanwu Center for Fundamental Theory, Xi'an 710127, China
\and Fundamental Discipline Research Center for  Quantum Science and technology of Shaanxi Province, Xi'an 710127, China}

\date{\today}

\abstract{
Studies of QCD phase transition and critical signals are often conducted under spatially uniform temperature conditions. The influence of spatially varying temperature on the phase transition signals in the fireball at the collider has not yet been fully explored. Based on an Ising-like effective potential, we study the locally equilibrated systems with spatially non-uniform temperature background. In a 2D disk geometry, the low-energy fluctuation spectrum is explicitly resolved into radial and angular momentum modes. The nonlocal correaltions of single eigen-mode exhibits strong anisotropy, which are long-ranged along isotherms but suppressed radially due to the thermal geometry of the system. Unlike homogeneous systems where the zero-momentum mode dominates, correlations in such inhomogeneous system result from the superposition of a series of zero and non-zero angular momentum modes with comparable contributions in the vicinity of the phase transition interface. We extract the angular momentum modes and establish their connection to experimentally observable anisotropic flow. We find azimuthally sensitive observables may offer a previously unexplored avenue for detecting the QCD phase transition.
}


\maketitle

\section{Introduction\label{sec:intro}}
A central goal in high-energy nuclear physics is to understand the QCD phase transition under various conditions of temperature and baryon density \cite{STAR2010,STAR:2017sal,Luo:2017faz,Bzdak:2019pkr}. Significant theoretical progress has been made in mapping the QCD phase diagram using lattice QCD, effective models, and functional methods \cite{Brown:1990,Aoki:2006we,Aoki:2009sc,Fischer:2018sdj,Roberts:2000aa,Qin:2011,Fischer:2014ata,Gunkel:2021oya,Gao:2020qsj,Gao:2020fbl,Berges:2000ew,Dupuis:2020fhh,Fu:2022gou,Fu:2019hdw,Herbst:2011,Klevansky:1992qe,Fukushima:2004qe,Fu:2007xc,Jiang:2013,Schaefer:2007,Schaefer:20072,Shao:2011fk,He:2013qq,Chelabi:2015gpc,Chen:2020ath}. Experimentally, the hot and dense quark-gluon plasma (QGP) produced in relativistic heavy-ion collisions provides a unique laboratory for probing the QCD phase structure through its dynamical evolution \cite{TDLee:1974kw,Gyulassy:2004zy,BRAHMS:2004adc,STAR:2005gfr,Song:2010mg,Busza2018}. Critical signals are invoked to account for deviations between experimental measurements and non-critical background expectations, as observed in quantities such as net-proton fluctuations \cite{Stephanov:2008qz,Athanasiou:2010kw,Stephanov:2011pb,Fu:2021oaw,STAR:2020tga}, light nuclei yield ratios \cite{Sun:2018jhg,STAR:2022hbp}, baryon-strangeness correlations \cite{Koch:2005vg,Bollweg:2024epj}, and intermittency in charged hadron distributions \cite{Wu:2019mqq,STAR:2023jpm}.

The interpretation of these deviation signals relies heavily on theoretical modeling of critical fluctuations. Critical dynamics are governed by the slow (i.e., low-energy) modes of the QCD order parameter field (denoted as the $\sigma$ field).  These modes lie beyond the scope of standard hydrodynamics which assumes local thermal equilibrium with rapid relaxation. Under the assumption of isothermal conditions, studies of the zero mode of the $\sigma$ field predict non-monotonic behavior in higher-order cumulants \cite{Stephanov:2008qz}. Extensive efforts have been devoted to understanding non-equilibrium dynamics and finite-size effects on the evolution of fluctuations, through frameworks such as Langevin and Fokker-Planck equations \cite{Berdnikov:1999ph,Stephanov:2009ra,Mukherjee:2015swa,Jiang:2017mji,Jiang:2021zla,Sakaida:2017rtj,Nahrgang:2018afz,Jiang:2023nmd}. More recently, modern hydrodynamic models have also incorporated isothermal critical correlators to provide a more realistic dynamical description that includes contributions from critical modes \cite{Nahrgang:2011mg,Stephanov:2017ghc,Du:2020bxp,An:2020vri}.

However, the isothermal approximation, which is commonly employed in studies of critical modes in a local finite system, neglects the global influence of a spatially non-uniform background on the slow modes and their associated signals. In a previous study, we demonstrated that such non-uniform temperature distribution can qualitatively alter the physics of the phase transition in a one-dimensional geometry \cite{Zheng:2021pia}. In this work, we extend that investigation to a more realistic setting by studying critical fluctuations of the $\sigma$ field in a  steady and inhomogeneous thermal background with cylindrical symmetry.

In Section~\ref{sec2}, we formulate the probability distribution functional of the $\sigma$ field using an Ising-like effective potential and a Gubser flow-inspired temperature profile, and derive the base order parameter field which maximizes the probability distribution in the presence of a stationary non-uniform temperature background. In Section~\ref{sec3}, we analyze the linearized fluctuations around this base field and compute the spectrum and spatial profiles of the resulting excited modes. These low-energy modes are localized near the phase-transition interface and are classified by their angular momentum quantum number  owing to the rotation symmetry of the system. The system consequently displays intrinsically anisotropic spatial correlations, which are characterized by long-range coherence along isotherms and strong suppression in the radial direction.
Decomposing the fluctuations enables a direct analysis of how the thermal geometry redistributes fluctuation strength across different modes. In contrast to a homogeneous system, the low-energy spectrum in the inhomogeneous background always exhibits a nonzero spectral gap. Moreover, the combined effect of the eigenmodes with distinct angular momentum quantum numbers reduces the correlation length.
In Section~\ref{sec4}, we extract distinct angular momentum eigenmodes of the fluctuating $\sigma$ field  and examine the correlation structure of each component. The modes with nonzero angular momentum show a direct correspondence to anisotropic flow harmonics, which are well-established observables in heavy-ion collision experiments. We conclude in Section~\ref{sec5} with a summary of our main results and an outlook on future work.

\section{The theoretical framework} \label{sec2}

\subsection{The partition function}

To qualitatively capture key features of the fireball system and simplify the modeling, the following assumptions are adopted:

First, since this work focuses on critical fluctuations during the phase transition, we treat the slow modes of the $\sigma$ field as the sole relevant degree of freedom. All the other degrees of freedom such as quarks, gluons, hadrons, and the fast modes of the $\sigma$ field, are assumed to relax rapidly into local thermal equilibrium and thereby constitute a thermal background that determines the spatially varying temperature and baryon chemical potential profile. The effective Hamiltonian density governing the $\sigma$ field is then given by
\begin{equation}
    \mathcal{H}(\hat\pi,\hat\sigma) = \hat\pi^{2}/2 + (\nabla \hat\sigma)^{2}/2+\mathcal{V}(\hat\sigma,T,\mu),
\end{equation}
where $\hat\pi(\bm r)$ is the conjugate momentum for the order parameter field $\hat\sigma(\bm r)$, and the effective potential $\mathcal{V}(\hat\sigma,T,\mu)$  incorporates contributions from all integrated-out degrees of freedom.

Second, the majority of the fireball lies outside the critical region (i.e., away from the phase-transition interface) where critical slowing down effects are weak. In these noncritical regions, the $\sigma$ field relaxes rapidly into local thermal equilibrium. Only within the narrow critical region, the $\sigma$ field experiences critical slowing down and is subject to boundary constraints imposed by the surrounding equilibrated $\sigma$ field in the noncritical regions.

Third, since our primary goal is to investigate how temperature inhomogeneity influences critical fluctuations, we assume that the background chemical potential is spatially uniform and the temperature profile remains approximately static over the relevant timescale. This allows us to neglect real-time dynamical evolution and treat the system as being in a quasi-steady state. Under this assumption, the $\sigma$ field is taken to be in instantaneous thermal equilibrium with the prescribed thermodynamic background.

Finally, we adopt the local equilibrium ansatz for the statistical weight of $\sigma$ field configurations. The probability of a given configuration is governed by a spatially varying Boltzmann factor constructed from the local temperature. Specifically, the contribution to the partition function from a small spatial volume element  $\Delta \bm r$ centered at $\bm r$  is approximated as
\begin{equation}\label{eq:2}
\mathcal{Z}_{\Delta\bm r} = \text{Tr} \exp \left\{-\int_{\Delta\bm r} d^3\bm r \frac{\mathcal{H}(\hat \pi,\hat\sigma)}{T(\bm r)}
\right\},
\end{equation}
where the trace ``{Tr}'' denotes the functional integral over all field configurations within $\Delta \bm r$, and  $\hbar = k_B =1$ is kept throughout the paper. This expression assumes a locally equilibrated thermal background and neglects the effects of the background fluid velocity. The full partition function for the order parameter field is then formally expressed as the product over all spatial cells:
\begin{equation} \label{eq:3}
\mathcal{Z} = \prod_{\bm r} \mathcal{Z}_{\Delta\bm r}=\text{Tr} \exp \left\{-\int d^3\bm r \frac{\mathcal{H}(\hat\pi,\hat\sigma)}{T(\bm r)}
\right\}.
\end{equation}
Because the temperature $T(\bm r)$ varies spatially, it cannot be factored out of the integral. Instead, the combination defines an effective Hamiltonian density $\mathcal{H}_{\text{eff}}(\pi,\sigma) = \mathcal{H}(\pi,\sigma)/T(\bm{r})$ and a corresponding spatially uniform effective temperature $T_{\text{eff}}=1$ (dimensionless).

Write the partition function in the path-integral form
\begin{equation}
\mathcal{Z} = \int \mathcal{D}\pi \int_{\text{periodic}}\mathcal{D}\sigma  \exp\left\{ \int_0^1 d\tau  \int d^3\bm r \left[i\pi \partial_\tau \sigma - \mathcal{H}_{\text{eff}}(\pi,\sigma) \right]
\right\},
\end{equation}
where `periodic' means that the $\sigma$ field satisfies $\sigma(\bm r, \tau=0) = \sigma(\bm r, \tau=1)$. Integrating out the $\pi$ field gives
\begin{equation} \label{eq:action0}
\mathcal{Z}=  \int_{\text{periodic}}\mathcal{D}\sigma e^{S[\sigma]}.
\end{equation}
and the action has the form
\begin{equation}\label{eq:action}
S[\sigma]= -\int_0^1 d\tau  \int d^3\bm r \left[  \frac{T  (\partial_\tau \sigma)^2}{2} + \frac{(\nabla
\sigma)^{2}}{2T}  + \frac{\mathcal{V}(\sigma,T,\mu)}{T}\right] .
\end{equation}

The $\sigma$ field maximizing the weight $e^S$ in the partition function is named the base $\sigma$ field (i.e., the saddle-point approximation) and denoted by $\sigma_c$. The $\sigma_c$ field satisfies the extremum condition
\begin{equation}
\left.\frac{\delta S}{\delta \sigma }\right|_{\sigma =\sigma_c}=0,
\end{equation}
or explicitly,
\begin{equation}\label{eom1}
-T \partial_\tau^{2}\sigma - \frac{1}{T}\nabla ^{2}\sigma  + \frac{\nabla \sigma \cdot \nabla T}{T^2}+ \frac{1}{T}\frac{\partial }{\partial \sigma }\mathcal{V}(\sigma,T,\mu) =0.
\end{equation}
Here, we have adopted the uniform effective temperature $T_{\text{eff}}=1$. The cost of this treatment is that the derivatives acquire a metric regulated by a position-dependent temperature. The corresponding effective Matsubara frequencies of $\mathcal{H}_{\text{eff}}(\pi,\sigma)$ are uniform throughout space, with the temperature acting as a prefactor. In Section \ref{sec3.1}, we will further discuss the effect of non-uniform temperature on the thermal fluctuations.

\begin{figure*}[tbp]
\centering
\includegraphics[width=1.65\columnwidth]{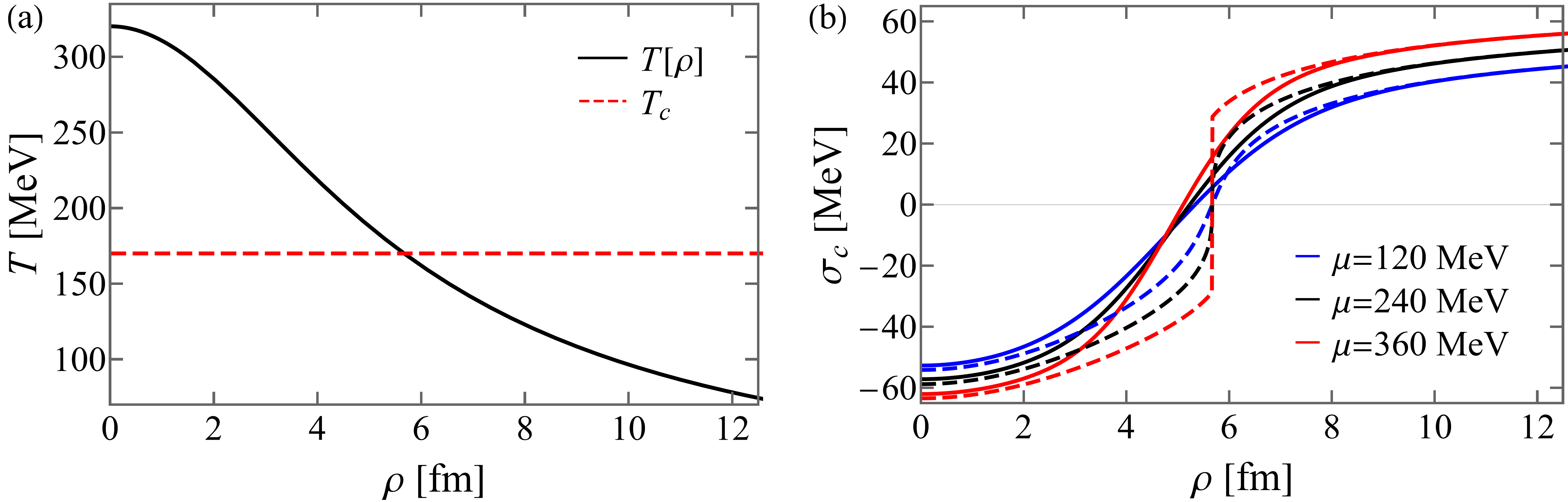}
\caption{ (a) The spatial temperature distribution as a function of the radius $\rho$. (b) The base field $\sigma_c$ as a function of $\rho$. The solid (dashed) lines are solutions to Eq. (\ref{eom}) at different chemical potentials, and for comparison, the dashed lines are results determined by ${\partial_\sigma }\mathcal{V} = 0$. }
\label{fig1}
\end{figure*}

%
%
%

\subsection{The base $\sigma$ field}

The base configuration $\sigma_c$, which minimizes the effective action, must satisfy $\partial_\tau \sigma_c = 0$ from equation \eqref{eq:action}. In principle, the effective potential $\mathcal{V}(\sigma,T,\mu)$ should be derived from phenomenological models. However, since the QCD phase transition belongs to the same universality class as the 3D Ising model, we employ a parameterized Ising-like potential for a universal description and computational tractability. Specifically, we adopt a similar form to that in \cite{Zheng:2021pia,Jiang:2023nmd}
\begin{equation}\label{veff1}
\mathcal{V}(\sigma,T,\mu) = \eta_1 \sigma + \eta_2\sigma^2/2 + \eta_4\sigma^4/4.
\end{equation}
The coefficients in the potential are assumed to be the following forms:
\begin{eqnarray}
  \eta_1 &=& 0.5 \text{fm}^{-2} \times (T-T_c), \label{veff2} \\
  \eta_2 &=& -0.5 \text{fm}^{-1}  \times (\mu-\mu_c), \label{veff3} \\
  \eta_4 &=& 14.4. \label{veff4}
\end{eqnarray}
With these parameters, the potential can exhibit different phase transition scenarios when decreasing temperature, like crossover at small $\mu$, and first-order phase transition at large $\mu$. The location of the critical point is also an input, and is set to be at $(T_c, \mu_c) = (170, 240)$\text{MeV}.

In hydrodynamic analysis, under the boost-invariance approximation, the 3+1 dimensional problem reduces to 2+1 dimensions. In the local longitudinal comoving frame of each rapidity slice, the system evolves independently in each slice but follows the same dynamical equations. Thus, only transverse expansion, driven by the nonuniform temperature and chemical potential profiles, needs to be considered. We apply the local equilibrium ansatz to each rapidity slice. In the comoving frame, the transverse flow velocity remains finite; however, its effects on the critical fluctuations have been neglected in our ansatz (equation \eqref{eq:3}) for simplicity.
For central collisions, we further assume rotational symmetry in the transverse plane. Thus, at any fixed spacetime rapidity, the system simplifies to a disk of thickness $d_z$ (in the comoving frame) in the longitudinal direction, and the temperature depends only on the radial position, $T(\mathbf{r}) = T(\rho)$, where $\rho$ is the radial distance from the center of the disk.
With the cylindrical coordinates $(\rho,\theta, z)$ and the zero-momentum-mode approximation in $z$-direction, the above equation (\ref{eom1}) becomes
\begin{equation}
-\frac{1}{\rho}\frac{\partial }{\partial \rho}\left( \rho\frac{\partial \sigma
}{\partial \rho}\right) + \frac{\hat{L}^2_z}{\rho^{2}}\sigma + \frac{1
}{T}\frac{\partial \sigma }{\partial \rho}\frac{\partial T}{\partial \rho}+\eta_1+\eta_2\sigma +\eta_4\sigma ^{3} = 0,
\end{equation}%
where the angular momentum operator reads
\begin{equation}
\hat{L}_z = -i\frac{\partial}{\partial
\theta}.
\end{equation}
For the base field, the angular momentum is zero. As a result, the extremum
condition turns to
\begin{equation}\label{eom}
-\frac{\partial ^{2}\sigma _{c}}{\partial \rho^{2}} - \frac{1}{\rho}\frac{\partial
\sigma _{c}}{\partial \rho} +\frac{\partial \sigma _{c}}{\partial \rho}\frac{
\partial \ln T}{\partial \rho}+ \eta_1+\eta_2\sigma_c +\eta_4\sigma_c ^{3} = 0.
\end{equation}
The temperature profile of the system is assumed to be
\begin{equation}\label{temp}
T(\rho) = \frac{C}{t}\frac{(2qt)^{2/3}}{\left[1+2q^2(t^2+\rho^2)+q^4(t^2-\rho^2)^2\right]^{1/3}},
\end{equation}
the form of which is taken from the Gubser temperature solution \cite{Du:2020bxp,Gubser:2010ze}. We emphasize that this form is adopted only as a phenomenological representation of an inhomogeneous thermal background. Such a temperature profile would generally induce a non-zero fluid velocity field through the hydrodynamic equations; however, the present work is based on a quasi-steady-state analysis, and the effects of fluid velocity are not included. In this respect, our current setup does not provide a full treatment of a self-consistent dynamical framework, a task we leave for future work. For simplicity, the baryon chemical potential is assumed spatially uniform in the following analysis. As we mainly focus on the instantly steady distribution of the $\sigma$ field, we choose a temperature profile at a specific time point by setting $t = 1$ fm, and the other parameters are $C=2.8$, and $q={1}/{4.3}~\text{fm}$. The spatial dependent temperature is plotted in Figure \ref{fig1} (a), which shows that the temperature decreases as the increase of radius $\rho$.

To illustrate the impact of the inhomogeneous temperature on the base field, we numerically solve equation (\ref{eom}). The resulting base field configurations $\sigma_c(\rho)$ (solid lines) for $\mu = 120, 240, 360$ MeV are shown in Figure \ref{fig1} (b). For comparison, dashed lines show the results with local equilibrium approximation, obtained by pointwise minimization of the effective potential, ${\partial_\sigma }\mathcal{V}(\sigma,T(\rho),\mu) =0$, which neglects spatial gradients and thus ignores the nonlocal constraints imposed by equation (\ref{eom}) across space. Away from the phase transition region (or critical region), the base field $\sigma_c$  coincides with the local equilibrium value, which demonstrates that the $\sigma$ field in the critical region is indeed constrained by the surrounding equilibrated $\sigma_c$ field in noncritical regions.

The phase transition is identified by the sign change of  the base field $\sigma_c$. In the local equilibrium approximation, the transition occurs at the same temperature $T = 170 $MeV (corresponding to $\rho = 5.7$ fm) for all three cases, as dictated by the parameterization of the effective potential \eqref{veff1}-\eqref{veff4}. However, when spatial gradient terms are incorporated, the $\sigma$ field becomes smooth across the transition region, even in the first-order scenario at $\mu = 360$ MeV. Moreover, the location of the sign change of $\sigma_c$ (representing the effective transition point indicated by the $\sigma$ field) shifts to smaller radii where the background temperature is higher. This shift becomes more pronounced at larger chemical potentials.

\section{Fluctuations and correlations} \label{sec3}
\subsection{Eigen-mode fluctuations} \label{sec3.1}

In the following, we will deduce the fluctuations and correlations for such inhomogeneous system by pertubatively expanding the probability distribution \eqref{eq:action0} around the base field $\sigma (\bm r, \tau) =\sigma _{c}(\bm r) + \tilde\sigma (\bm r, \tau)$, with $\tilde\sigma (\bm r, \tau)$ representing the fluctuations.
Note that the fluctuation term depends both on $\rho$ and the azimuths $\theta$. The action is divided into two parts: $ S[ \sigma _{c}(\bm r)]$ and the fluctuating part  $\Delta S[ \tilde\sigma ( \bm r,\tau ) ]$,
\begin{equation} \label{part}
 \Delta S[\tilde\sigma ( \bm r, \tau) ]  = -\int_0^1 d\tau  \int d^3\bm r  \left(\frac{1}{2}\tilde\sigma \hat{\mathcal{O}}\tilde\sigma
 +\frac{\lambda_3}{3} \tilde\sigma^{3}+ \frac{\lambda_4}{4} \tilde\sigma^{4}
 \right),
\end{equation}%
where
\begin{eqnarray}
\hat{\mathcal{O}}& = & \overleftarrow{\partial_\tau}
T  \overrightarrow{\partial_\tau} + \overleftarrow{\nabla}
\frac{1}{T} \cdot \overrightarrow{\nabla} + V_{\text{eff}}(\rho),  \\
V_{\text{eff}}(\rho) &=& \frac{ ( \eta_2+3\eta_4\sigma _{c}^{2}) }{T},  \\
\lambda_3  & = & \frac{3\eta_4 \sigma _{c}}{T}, \\
\lambda_4 &=& \frac{\eta_4}{T}.
\end{eqnarray}%
The arrows on the hatted operators indicate the direction the operators act along.

In the perturbation theory, the non-Gaussian fluctuations are treated as high-order
terms, and will be calculated in the later section. For the Gaussian part, the
fluctuating $\tilde\sigma(\bm r,\tau)$ can be expanded in terms of the orthogonal, complete set of eigenfunctions of the Hermitian operator $\hat{\mathcal{O}}$. It can be verified that $\hat{L}_z$ commutes with $\hat{\mathcal{O}}$, since $T$, $\eta_{1,2,4}$ and $\sigma_c$ are rotationally symmetric about the $z$-axis. Therefore, under the zero-momentum-mode approximation along the $z$ direction, $\tilde\sigma$ is expanded as
\begin{equation}
\tilde{\sigma}\left( \bm{r},\tau \right) =\sum_{\lambda,l=-\infty }^{\infty}\sum_{n}c_{\lambda nl}R_{\lambda nl}\left( \rho\right) \Theta _{l}\left( \theta \right)e^{-i\omega _{\lambda }\tau },
\end{equation}
where the Matsubara frequency is $\omega _{\lambda}=2\pi \lambda$ with $\lambda \in \mathbb{Z}$, owing to the periodicity condition $\sigma \left( \bm r,0\right) =\sigma \left(\bm r, 1\right) $. $\Theta _{l}\left( \theta \right) = e^{il\theta }/{\sqrt{2\pi}}$, with $l$ denoting the angular momentum quantum number. $n$ is the integer index labeling the radial excitation level for given $\lambda$ and $l$. The radial and angular eigenfunctions are orthonormalized separately as:
\begin{eqnarray}
\int_{0}^{\infty } R_{\lambda n^{\prime }l}^*\left(
\rho\right) R_{\lambda nl}\left( \rho\right) \rho d\rho &=&\delta _{nn^{\prime }}, \label{eq:23} \\
\int_{0}^{2\pi }\Theta _{l^{\prime }}^{\ast }\left( \theta \right) \Theta
_{l}\left( \theta \right) d\theta  &=&\delta _{l^{\prime }l}.
\end{eqnarray}%
The corresponding eigenvalue of the operator $\hat{\mathcal{O}}$ for the eigenfunction $R_{\lambda nl}\left( \rho\right) \Theta _{l}\left( \theta \right)e^{-i\omega _{\lambda }\tau }$ is denoted by $E_{\lambda nl}$. Since $\sigma$ is real, $\tilde{\sigma} = \tilde{\sigma}^*$. Thus, the quadratic part of the action becomes
\begin{equation}\label{actionx}
\Delta S[ \tilde\sigma] = -\frac{1}{2} \int_0^1 d\tau  \int d^3\bm r \tilde\sigma^* \hat{\mathcal{O}}\tilde\sigma = -\frac{1}{2} d_z\sum_{\lambda nl} c^{*}_{\lambda  nl} E_{\lambda nl} c_{\lambda nl} ,
\end{equation}
where $d_z$ is the (constant) thickness of the disk.
Since the normalization in equation \eqref{eq:23} involves the non-trivial measure \(\rho d\rho\), it is convenient to introduce a complete basis in the radial direction, \(\sqrt{\rho}\,R_{\lambda n l}(\rho)\), so that the orthonormality condition becomes \(\int_{0}^{\infty} [\sqrt{\rho}R_{\lambda n'l}]^* [\sqrt{\rho}R_{\lambda nl}] \, d\rho = \delta_{nn'}\), with the radial measure \(d\rho\). Correspondingly, \(\Delta S[\tilde\sigma]\) can be written in the following form:

\begin{equation}
\Delta S[ \tilde\sigma]
= -\frac{1}{2} d_z \sum_{\lambda nn'l} c^*_{\lambda n'l}  c_{\lambda  nl} \int d \rho [\sqrt{\rho} R^*_{\lambda n'l}] \hat{\mathcal{O}}_{\lambda  l}  [\sqrt{\rho} R_{\lambda nl}], \\
\end{equation}
where
\begin{eqnarray}\label{onl}
   \hat{\mathcal{O}}_{\lambda l} &=& T(\rho) \omega_\lambda ^2 + \overleftarrow{\partial_\rho} \frac{1}{T(\rho)} \overrightarrow{\partial_\rho}- \overleftarrow{\partial_\rho}\frac{1}{2\rho}\frac{1}{T(\rho)}- \frac{1}{2\rho}\frac{1}{T(\rho)} \overrightarrow{\partial_\rho} \notag \\
    && + \frac{1}{4\rho^{2}}\frac{1}{T\left( \rho\right) }+\frac{1}{T\left(\rho\right) }\frac{l^{2}}{\rho^{2}}
   + V_{\text{eff}}(\rho),
\end{eqnarray}
is a Hermitian operator.  Comparing with equation \eqref{actionx}, we can see that \(\sqrt{\rho}R_{\lambda nl}(\rho)\) is indeed the eigenfunction of \(\hat{\mathcal{O}}_{\lambda l}\) with eigenvalue \(E_{\lambda nl}\).  In the numerical calculation, we obtain \(E_{\lambda nl}\) and \(\sqrt{\rho} R_{\lambda nl}\) directly from the Hermitian operator \(\hat{\mathcal{O}}_{\lambda l}\), rather than from \(\hat{\mathcal{O}}\).

\begin{figure*}
\centering
\includegraphics[width=1.65\columnwidth]{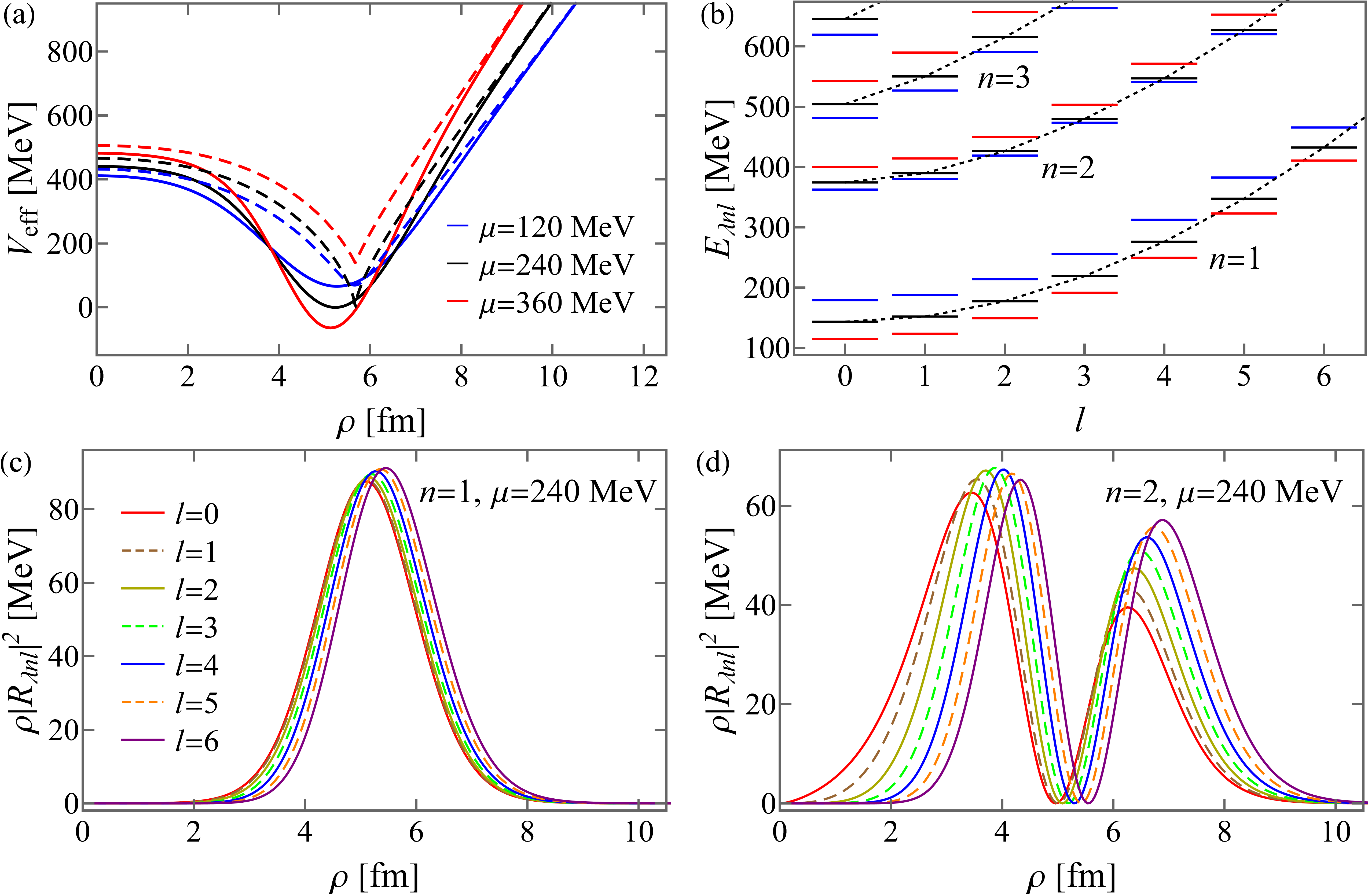}
\caption{(a) The radial dependence of the effective potential $V_{\text{eff}}$.
The solid lines correspond to $\sigma_c$ in $V_{\text{eff}}$ determined by equation\,\eqref{eom}, while dashed lines are results with the profile obtained from the condition ${\partial_\sigma }\mathcal{V} = 0$. Different colors are for results with three chemical potential values: $\mu=120$ MeV (blue), $240$ MeV (black), and $360$ MeV (red). (b) The corresponding spectra of fluctuations for different radial ($n$) and angular momentum ($l$) modes, using the same color scheme for $\mu$. Panels (c) and (d) display the radial probability densities for the first ($n=1$) and second ($n=2$) radial excitations, respectively, with different colors denoting different angular momentum modes and the chemical potential fixs at $\mu=240$ MeV. For all panels, the Matsubara mode is $\lambda=0$. }
\label{figure2}
\end{figure*}

In Figure \ref{figure2} (a), we display the effective potential $V_{\text{eff}}(\rho)$ for fluctuations, calculated using the base configuration $\sigma_c$ determined either by the extremum condition \eqref{eom} (solid lines) or by the local equilibrium condition ${\partial_\sigma }\mathcal{V} = 0$ (dashed lines). In the local equilibrium condition, $V_{\text{eff}}(\rho)$ corresponds to the local effective mass squared of the $\sigma$ field (in unit of temperature $T$), which is always non-negative and vanishes only at the critical point. In contrast, the solid curves obtained from the extremum condition exhibit a pronounced softening and broadening in the phase transition region ($\rho \approx 4-7$ fm) for all three chemical potentials $\mu=120$, $240$, $360$ MeV. Notably, $V_{\text{eff}}(\rho)$ can be negative in the first-order phase transition region ($\mu >240 $ MeV). This modification of the effective potential arises from the spatially non-uniform temperature profile, which alters the base field configuration and, consequently, reshapes the effective potential near the critical interface.

By setting the disk radius to $20$ fm, we discretize the radial coordinate into 480 grid points and construct a finite-difference matrix representation of the operator $\hat{\mathcal{O}}_{\lambda l}$ defined in equation (\ref{onl}). Numerical diagonalization of this matrix yields the spectra $E_{\lambda nl}$ and the corresponding eigenfunctions $R_{\lambda nl}(\rho)$ for quantum fluctuations governed by the effective potential. Figure \ref{figure2} (b) shows the fluctuation spectra $E_{\lambda nl}$ for the zero Matsubara mode ($\lambda=0$) as a function of the radial quantum number ($n$) and angular momentum quantum numbers ($l$). All eigenvalues are strictly positive, confirming the stability of the base configuration, even though the effective potential itself can become negative in the first-order phase transition region.
Importantly, the lowest excitation mode possesses a finite energy value (energy gap) for all values of the chemical potential, in stark contrast to the zero-energy Goldstone mode expected at the critical point in a homogeneous system with translational symmetry. Specifically, for $n=1$ and $\lambda = 0$, the spectrum approximately has the form $E_{\lambda nl} = \Delta + \alpha l^2$, where $\Delta$ is the energy gap and $\alpha$ can be obtained by fitting the $E_{\lambda nl}\sim l$ curve. For the current parameter setting, $\Delta = 181.85, 145.12, 116.14$ MeV and $\alpha = 7.97, 8.04, 8.23$ MeV for $\mu = 120, 240, 360$ MeV, respectively. The gap in the first-order region is even smaller than that at the critical point.

For a fixed radial quantum number $n$, the eigenvalues increase with angular momentum $l$. This trend originates from the centrifugal potential term $l^2/(\rho^2 T)$ in the operator (\ref{onl}), which energetically disfavors azimuthal fluctuations. The resulting spectrum illustrates how temperature inhomogeneity restructures the low-energy fluctuation modes. Importantly, applying an energy cutoff near $450$ MeV selects only modes with $n=1, l \lesssim 6$ and $n=2, l \lesssim 2$. Lowering the cutoff to $250$ MeV further restricts the spectrum to modes with $n=1$ only. This indicates that radial excitations are suppressed at lower energy scale, leaving a series of azimuthal fluctuations as the dominant low energy degrees of freedom near the critical point.

Figure \ref{figure2} (c) and (d) show the profiles of $\rho |R_{\lambda nl}(\rho)|^2$ for the $n=1$ and $n=2$ radial excitations, respectively, at $\mu=240$ MeV. The colored lines correspond to different angular momentum quantum numbers $l$. For the $n=1$ modes in panel (c), a dominant feature is the strong spatial localization of all fluctuations within the phase transition region ($4 \lesssim \rho \lesssim 7$ fm), which coincides with the region of effective potential softening shown in panel (a). This confirms that critical fluctuations are not homogeneously distributed but are instead confined to the narrow interfacial region where the local temperature crosses the critical value $T_c$. Moreover, increasing $l$ slightly shifts the peak of the profile outward to larger radii, which is a direct consequence of the centrifugal potential. For the $n=2$ modes in panel (d), the profile of $\rho |R_{\lambda nl}(\rho)|^2$ exhibits a single radial node in $R_{\lambda nl}(\rho)$ as expected. Similarly, higher values of $l$ induce a modest outward shift of the eigenfunction.

\begin{figure*}[tbp]
\centering
\includegraphics[width=1.7\columnwidth]{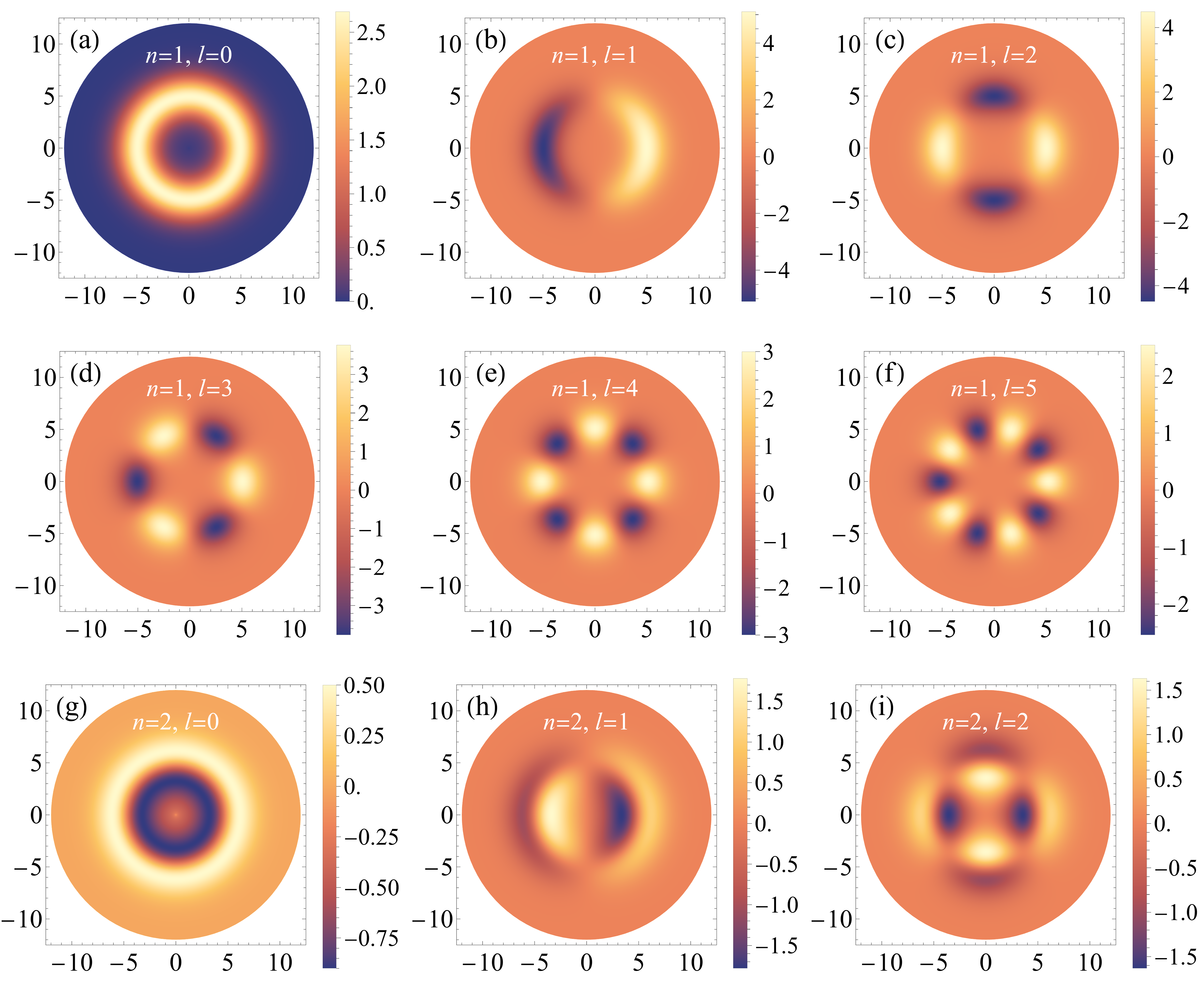}
\caption{Individual contributions of the first few low-energy fluctuation modes to the nonlocal two-point correlation function $\langle\tilde{\sigma}(\bm{r},\tau)\tilde{\sigma}(\bm{r}^{\prime},\tau)\rangle d_z$ (in units of MeV). Each panel shows the correlation pattern resulting from a specific combination of radial ($n$) and angular momentum ($l$) quantum numbers (note that the results are summed over the $\pm l$ modes for $l>0$). For all the subplots, the anchor point $\bm{r}^{\prime}$ is fixed at $(5.83, 0)$ fm, the Matsubara mode is $\lambda=0$, and the chemical potential $\mu = 240$ MeV. The spatial coordinates $x$ and $y$ are given in fm.}
\label{figure3}
\end{figure*}

In the analysis above, we have focused on the eigenvalues and eigenfunctions of $\hat{\mathcal{O}}_{\lambda l}$ corresponding to the zero Matsubara mode with $\lambda=0$. In principle, the computation of correlation functions and related physical observables requires a sum over all Matsubara frequencies. We have examined the nonzero modes $\lambda = \pm 1, \pm 2, \cdots$ and find that the eigenvalues of $\hat{\mathcal{O}}_{\lambda l}$ increase rapidly with $|\lambda|$. Moreover, the associated eigenfunctions are predominantly localized near the outer boundary of the disk as a consequence of the $T(\rho) \omega_\lambda ^2$ term (which reaches a minimum at the outer boundary where the temperature is lowest) in $\hat{\mathcal{O}}_{\lambda l}$, and thus lie far from the phase transition interface where critical fluctuations are most pronounced. Consequently, the contribution of these non-zero Matsubara modes to correlation in the critical region are negligible. Based on this observation, we restrict all subsequent calculations to the zero Matsubara mode. This is equivalent to neglecting the $\tau$-dependence of the fluctuations.

Collectively, the results shown in Figure \ref{figure2} demonstrate that temperature inhomogeneity gives rise to a spatially localized ``critical band'' near the phase transition interface. Within this region, fluctuations are governed not by the zero-momentum mode that dominates in homogeneous systems, but rather by a family of low-energy modes with the lowest radial excitation ($n=1$) and a range of finite azimuthal quantum numbers ($l$). This indicates that critical fluctuations in the inhomogeneous system are intrinsically anisotropic, and that an inwardly hot, outwardly cold temperature profile enhances azimuthal excitations.
In addition, from a dimensional perspective, the low-energy fluctuations are confined to the narrow transition layer near the phase transition interface, indicating the effective dimensionality of the phase transition is effectively reduced by one.

Now we comment on the introduction of a uniform effective temperature \(T_{\rm eff}=1\) in the non-uniform temperature system. Naively, in a non-uniform temperature system, the Matsubara frequencies would in principle become spatially dependent from equation \eqref{eq:2}. By introducing a uniform effective temperature \(T_{\rm eff}=1\) while encoding the local temperature \(T(\rho)\) in a spatially varying effective Hamiltonian, the eigenmodes (eigen-wavefunctions) are no longer spatially uniform, in contrast to the uniform-temperature case where the eigenfunctions are plane-wave-like. The temperature inhomogeneity is now encoded in the spatial profiles of the eigenfunctions. For the zero Matsubara mode, the eigen-wavefunction is found to be concentrated in the phase-transition region and decays away from it. For non-zero Matsubara modes, in contrast, the eigen-wavefunctions are localized near the outer boundary of the disk and have negligible support in the phase transition region. Since our analysis focuses on fluctuations near the phase transition region, retaining only the zero Matsubara mode is well justified for the observables of interest. We note that the validity of this truncation is further supported by the hierarchy of scales: the term \(T(\rho)\omega_\lambda^2\) in equation \eqref{onl} for the lowest non-zero Matsubara mode (\(\lambda=\pm1\)) gives a contribution of order \(4\pi^2 T_c\) to the effective potential, well above the critical-mode energies (\(\sim 100\)--\(450\) MeV) that we focus on.

\begin{figure*}[tbp]
\centering
\includegraphics[width=1.65\columnwidth]{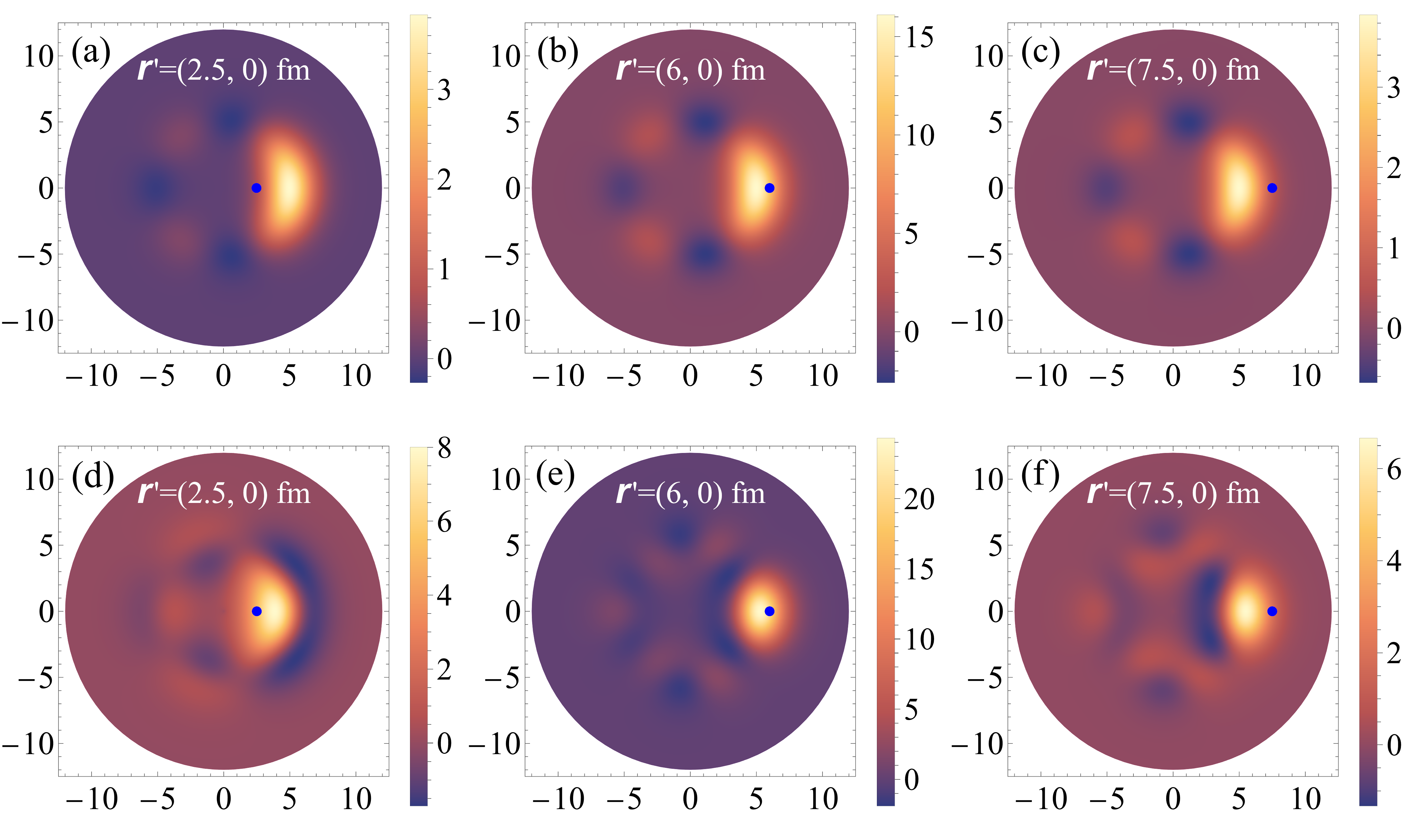}
\caption{The correlation $\langle\tilde\sigma({\bm r},\tau) \tilde\sigma({\bm r'},\tau)\rangle d_z$, in unit MeV, for the anchor point (marked as a blue point) fixed at $\bm r' = (2.5, 0)~\text{fm}$,  $ (6, 0)~\text{fm}$, and $ (7.5, 0)~\text{fm}$ respectively in each panel. In the first row, the summation are over low-energy modes below $250$ MeV, includes angular momentum  mode $l\leq 3$ for the first radial mode $n=1$. In the second row, the summation are over low-energy modes below $450$ MeV, includes angular momentum  mode $l\leq 6$ for $n=1$, and $l\leq 2$ for $n=2$. In all the subplots, the chemical potential $\mu = 240$ MeV and the Matsubara mode $\lambda=0$.}
\label{figure4}
\end{figure*}

\subsection{Spatial two-point correlation}

In this part, we investigate the spatial patterns of nonlocal two-point correlations arising from different
angular momentum channels, as well as the correlation structures resulting from their superposition. Since  $\sqrt{\rho}R_{\lambda nl}\left( \rho\right)$ and thus $R_{\lambda nl}$ are real functions and  the operator $\hat{\mathcal{O}}_{\lambda l}$ depends only on $\lambda^2$ and $l^2$, the function $R_{\lambda nl}\left(\rho\right) $ is independent of the signs of $\lambda$ and $l$.  Consequently,  $R_{\lambda nl}\left( \rho\right) =R_{-\lambda n-l}\left( \rho\right)$.
Using the fact that $\tilde\sigma$ is real, $\tilde{\sigma} = \tilde{\sigma}^*$, we find
\begin{equation}
    c_{-\lambda n-l}^{\ast }=c_{\lambda nl}.
\end{equation}
$c_{\lambda nl}$ is real for $\lambda =l=0$. From the action \eqref{actionx}, we obtain the correlation of the coefficients,
\begin{equation}
    \left\langle c_{\lambda 'n'l'}^{* }c_{\lambda nl}\right\rangle =\frac{1}{d_{z}E _{\lambda nl}}\delta _{\lambda '\lambda }\delta
_{n'n}\delta _{l'l}.
\end{equation}
As a result, the correlation of the fluctuating $\tilde\sigma$ becomes
\begin{equation}\label{correlation2}
    \langle\tilde\sigma({\bm r},\tau) \tilde\sigma({\bm r'},\tau) \rangle =\sum_{\lambda nl}\frac{R_{\lambda nl}\left( \rho \right) R_{\lambda nl}\left( \rho'\right)
\Theta _{l}\left( \theta \right) \Theta _{l}^{\ast }\left( \theta' \right) }{%
d_zE_{\lambda nl}}.
\end{equation}

Figure \ref{figure3} presents the individual contributions of the nonlocal two-point correlation function
$\langle\tilde{\sigma}(\bm{r},\tau)\tilde{\sigma}(\bm{r}^{\prime},\tau)\rangle d_z$ from first few low-energy eigen-modes ($E_{\lambda nl} \lesssim 450$ MeV).
The first two rows correspond to the lowest radial excitation ($n=1$) with increasing angular momentum mode, while the third row shows modes of the second radial excitation ($n=2$).
The specific ($n$, $l$) values are marked in each panel.
By fixing one point $\bm{r}^{\prime}$, named as the anchor point, in a location within the phase transition region ($\bm{r}^{\prime}=(5.83, 0)$ fm), its correlation with the other point is evaluated over the disk.
All results correspond to the chemical potential $\mu = 240$ MeV and the Matsubara mode $\lambda=0$.
For non-zero $l$, the plotted correlation represents the sum of the $+l$ and $-l$ modes.
In the numerical calculation, the system is defined on a disk of radius $20$ fm, and for clarity, only the inner region up to $10$ fm is displayed, since the relevant fluctuations are localized within this range.

Notably, for the $l=0$ mode, the correlation strength is uniform along any isotherm, regardless of the distance to the anchor point's location, which reflects the preservation of azimuthal (rotational) symmetry.
Higher-$l$ modes  exhibit nearly the same radial localization of correlation intensity as the $l =0$ mode but introduce distinct angular modulations. Each increment in $l$ generates a corresponding number of  alternating enhanced and suppressed correlation sectors along the isothermal ring, reflecting systematic angular quantization. From the third row, we observe that additional radial nodes appear as $n$ increases, whereas increasing $l$ introduces more nodes in the angular direction.


A similar observation in accord with the above section is that the nonlocal correlation are also spatially localized around the phase transition region ($\rho \approx 4-7$ fm) for different modes.
Moreover, the maximum correlation strength remains comparable across different
$l$ modes for a given $n$, indicating that non-zero angular momentum fluctuations carry weight comparable to the isotropic $l=0$ mode.
These results demonstrate that the global thermal geometry also plays a significant role in structuring the spatial nonlocal correlations. It not only localizes fluctuations radially but also organizes them into azimuthal harmonics, leading to anisotropic correlations along the radial and azimuthal direction.

\begin{figure*}
\centering
\includegraphics[width=1.75\columnwidth]{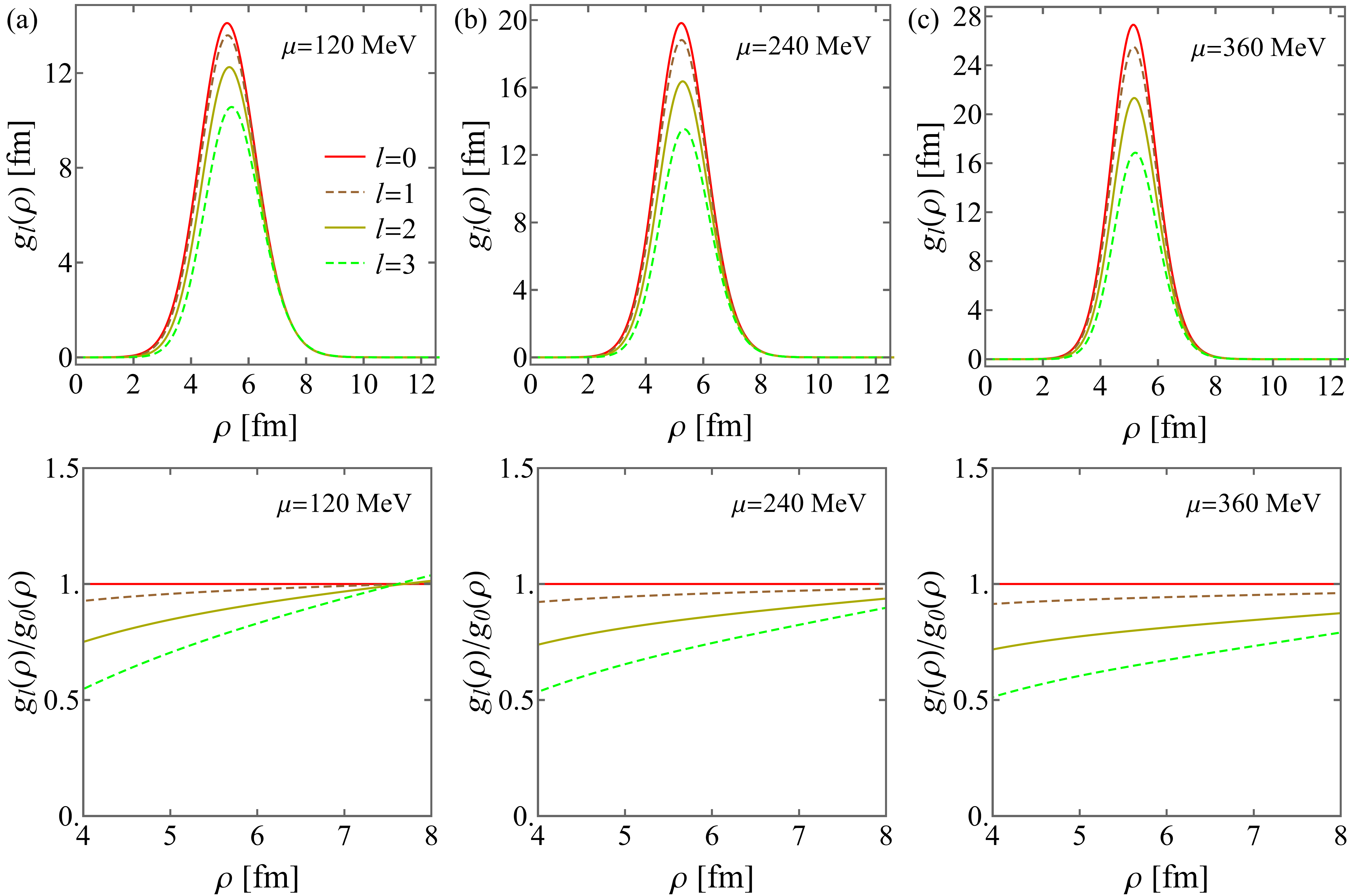}
\caption{Comparison of the two-point correlations $g_l(\rho) = \rho^2 \langle\tilde\sigma_{l}\left( \rho, \tau\right)\tilde \sigma_{-l}\left(
\rho, \tau\right)\rangle/d_z $ as function of $\rho$ for $l=0,1,2,3$ at $\mu=120$,$240$,and $360$ MeV, respectively.}
\label{figure5}
\end{figure*}

In Figure \ref{figure4}, we illustrate the cumulative effects of these distinct eigen-modes on the two-point correlation, with the anchor point $\bm r'$ (marked in blue) located at $(2.5, 0)$ fm, $(6, 0)$ fm, and $(7.5, 0)$ fm in each column, respectively. These points lie within the high-temperature, phase-transition, and low-temperature regions of the disk. The upper row shows results obtained by summing only modes with energy below $250$ MeV with quantum numbers $n=1, l\leq 3$. The lower row includes a broader set of modes below $450$ MeV with $n=1, l\leq 6$ and $n=2, l\leq 2$. The correlation profile is strongest when the anchor point resides inside the phase-transition region [panels (b) and (e)]. When the anchor lies outside this region, the overall correlation weakens, and its maximum shifts away from the anchor's original isotherm toward the phase-transition zone. This behavior follows directly from the structure of the radial eigenfunctions $R_{\lambda n l} (\rho)$ in equation \eqref{correlation2}, which are sharply peaked within the phase-transition region for the low radial mode.

A key observation is that summing over modes localizes the spatial correlation. The long-range azimuthal coherence of the $l=0$ mode is progressively fragmented by the inclusion of modes with nonzero angular momentum, gradually confining the dominant correlations to a localized region near the anchor point. At the same time, comparison between the two rows shows that by including more higher-energy modes, the correlation strengths are enhanced considerably near the anchor point. This conclusion is further confirmed when even more modes are involved. However, in practice, the high-energy modes are presumed to have been integrated out in our low energy effective model. Therefore, in the following calculations, we involve only the lowest branch ($n = 1$), which is localized in the phase transition region.


\section{Anisotropic correlations} \label{sec4}
From the above sections, we find the nonzero angular momentum modes make comparable contributions to the correlations. To observe the individual contributions of each angular momentum mode, we define a dimensionless quantity,
\begin{eqnarray}\label{sigmal}
\tilde{\sigma}_{l}\left( \rho,\tau \right)  &\equiv&  \int d z \int_{0}^{2\pi } d\theta \tilde{\sigma}%
\left( \bm r,\tau \right) e^{-il\theta } \notag \\
&=& {d_z\sqrt{2\pi }}\sum_{\lambda =-\infty}^{\infty }\sum_{n}c_{\lambda nl}R_{\lambda nl}\left( \rho\right)e^{-i\omega _{\lambda }\tau },
\end{eqnarray}%
in direct analogy to the experimental definition of anisotropic flow. Since the $\tilde{\sigma}$ field is real, one can verify from the above definition that $\tilde\sigma_{l}^*(\rho,\tau)= \tilde \sigma_{-l}(\rho,\tau)$.

\subsection{Two-point correlation}
The two-point correlation between different $l$-modes is
\begin{equation}
\left\langle \tilde{\sigma}^*_{l}\left( \rho,\tau \right) \tilde{\sigma}%
_{l'}\left( \rho',\tau'\right) \right\rangle  =
2\pi d_z \delta
_{l^{\prime }l}\sum_{\lambda =-\infty }^{\infty }G_{\lambda l}(\rho,\rho') e^{i\omega _{\lambda }\left( \tau -\tau'\right) },
\end{equation}%
where $G_{\lambda l}(\rho,\rho') =\sum_{n}{R_{\lambda nl}\left( \rho\right)
R_{\lambda nl}\left(\rho'\right) }/{E _{\lambda nl}}$. The correlation is nonzero only when $l'=l$, due to the rotation symmetry of the system. Note that the summation in the above equation is over all $\lambda$. However, in the critical region, the absolute dominant contribution comes from the $\lambda = 0$ term, as all other terms are negligible. Moreover, only the lowest branch ($n = 1$) is taken into account by assuming the effective energy scale of the system.

\begin{figure*}[t]
\large
\begin{center}
\includegraphics[width=1.7\columnwidth]{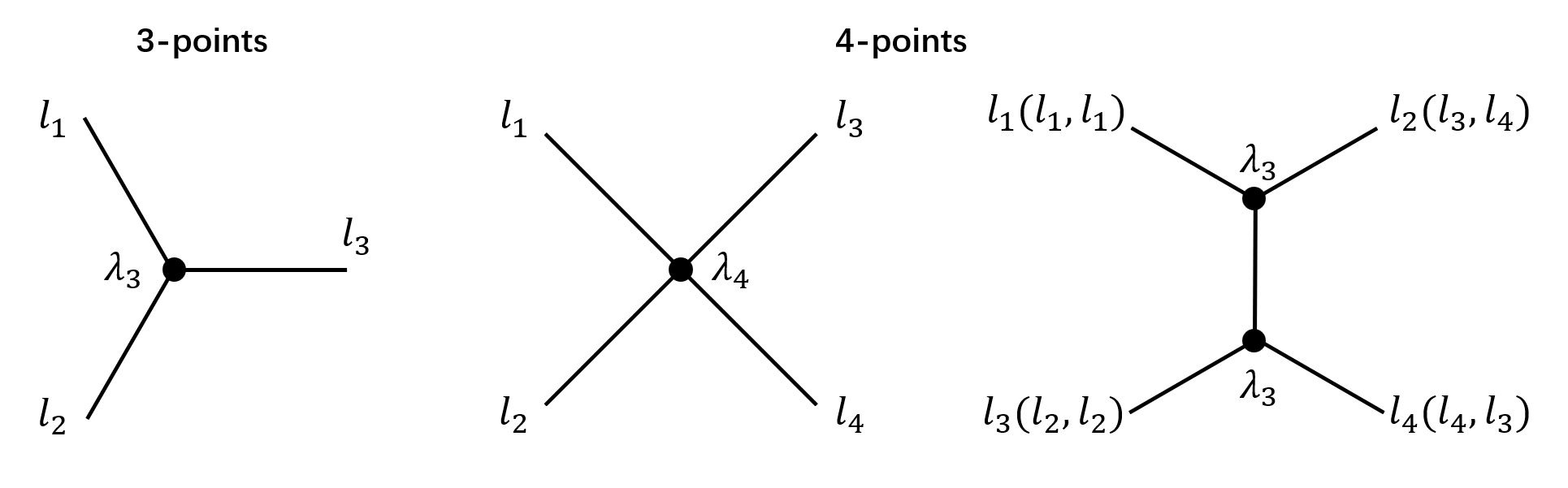}
\caption{Feynman diagrams for three-point correlation (left) and four-point correlation (center and right).}
\label{feynmann}
\end{center}
\end{figure*}

Figure \ref{figure5} presents the radial dependence of the local two-point correlation function $g_l(\rho) \equiv \rho^2 \langle\tilde\sigma_{l}\left( \rho, \tau\right)\tilde \sigma_{-l}\left( \rho, \tau \right)\rangle/d_z$ for different modes $l=0, 1, 2, 3$.  Different panels are for results with chemical potentials $\mu=120, 240,$ and $360$ MeV, respectively. A clear enhancement of correlations is observed in the phase transition region for different angular momentum modes.
The peak amplitude of $g_l(\rho)$ decreases with increasing $l$, consistent with the higher associated energy levels.
Notably, however, the strengths of anisotropic fluctuations are comparable among different modes, strongly different from homogeneous system where only the zero mode is dominant near the critical point. This is a direct consequence of a finite energy gap induced by the spatially non-uniform temperature distribution, which suppresses long-range coherence in the radial direction and redistributes fluctuation strength into finite angular momentum channels.

The same trend appears for all three chemical potentials, indicating that this fluctuation pattern is an intrinsic feature of systems with non-uniform temperature background, independent of the detailed order of the phase transition.
In such an inhomogeneous thermal background, fluctuations naturally populate multiple angular momentum channels.
Consequently, near-critical dynamics are fundamentally altered, a complete theoretical framework, such as Langevin or Fokker-Planck equations, must incorporate a series of finite-$l$ fluctuation modes in the treatments.

\subsection{Three-point and four-point correlation}

Based on the perturbation theory, we calculate the 3- and 4-point correlations for these nonzero angular momentum modes. The tree diagrams for the 3-point and 4-point correlations are plotted in Figure \ref{feynmann}.
After tedious reduction, the three-point equal-time correlation of different angular momentum modes is
\begin{eqnarray}
&&\Gamma_{l_1l_2l_3}^3(\rho) = \left\langle \tilde{\sigma}_{l_{1}}\left( \rho,\tau\right) \tilde{%
\sigma}_{l_{2}}\left( \rho,\tau\right) \tilde{\sigma}_{l_{3}}\left(
\rho,\tau\right) \right\rangle = -12\pi d_z \eta_4 \delta
_{l_{1}+l_{2}+l_{3},0} \notag \\
&& ~~~\times \sum_{\lambda _1,\lambda _2}  \int
 d\rho'\frac{\rho' \sigma _{c}(\rho')}{T\left(\rho'\right) }%
G_{\lambda_1l_1}(\rho,\rho')  G_{\lambda_2l_2}(\rho,\rho')
G_{\lambda_1+\lambda_2,l_3}(\rho,\rho'), \notag \\
\end{eqnarray}%
and the four-point equal-time correlation is
\begin{eqnarray}
&& \Gamma_{l_1l_2l_3l_4}^4(\rho) = \left\langle \tilde{\sigma}_{l_{1}}\left( \rho,\tau\right) \tilde{%
\sigma}_{l_{2}}\left( \rho,\tau\right) \tilde{\sigma}_{l_{3}}\left(
\rho,\tau\right)
\tilde{\sigma}_{l_{4}}\left( \rho,\tau\right) \right\rangle \notag\\
&& ~~~=  -12\pi d_z \eta_4 \delta
_{l_{1}+l_{2}+l_{3}+l_{4},0} \sum_{\lambda_1,\lambda_2,\lambda_3}  \int
 d\rho'\frac{\rho'}{T\left(\rho'\right) }%
\Big[G_{\lambda_1l_1}(\rho,\rho')  \notag\\
&& ~~~~~~~~~~ \times   G_{\lambda_2l_2}(\rho,\rho')   G_{\lambda_3l_3}(\rho,\rho')
G_{\lambda_1+\lambda_2+\lambda_3,l_4}(\rho,\rho') \Big] \notag\\
&&~~~~~~ +72\pi d_z\eta_4^2 \delta
_{l_{1}+l_{2}+l_{3}+l_{4},0}  \sum_{\lambda_1,\lambda_2,\lambda_3} \int{d\rho_1}\int{d\rho_2} \frac{\rho_1 \sigma_c(\rho_1)} {T\left(\rho_1\right) } \times  \notag\\
&&  ~~~~~~~~~~
\frac{\rho_2 \sigma_c(\rho_2)}  {T\left(\rho_2\right)}  \Big[G_{\lambda_1l_1}(\rho,\rho_1)  G_{\lambda_2l_2}(\rho,\rho_1)  G_{\lambda_3l_3}(\rho,\rho_2)   \notag\\
&&  ~~~~~~~~~~
\times G_{\lambda_1+\lambda_2+\lambda_3,l_4}(\rho,\rho_2)G_{\lambda_1+\lambda_2,l_1+l_2}(\rho_1,\rho_2)  \notag\\
&& ~~~~~~~~~~  + (l_2 \leftrightarrow l_3) + (l_2 \rightarrow l_4 \rightarrow l_3 \rightarrow l_2)
\Big].
\end{eqnarray} %
The Kronecker delta symbol appears in both the three- and four-point correlations due to the conservation of angular momentum. It acts as a selection rule among different $l$ modes, meaning that only specific combinations of $l$ values yield non-vanishing correlations.
In the low energy scale, even though the summation should be over all $\lambda$ and $n$ values, the dominant contribution comes from $\lambda = 0$ and $n = 1$.

\begin{figure*}
\centering
\includegraphics[width=2\columnwidth]{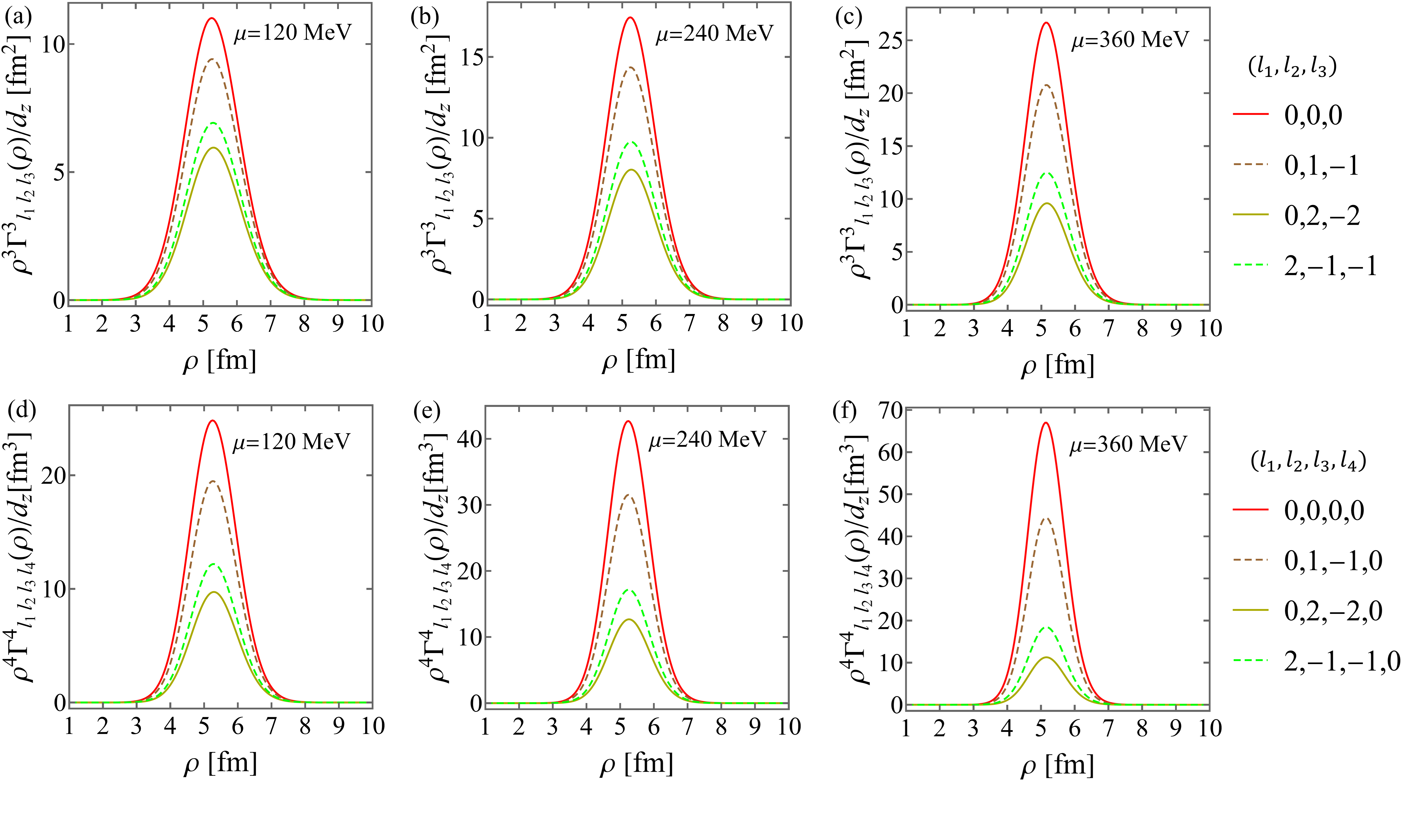}
\caption{The three-point (the upper row) and four-point correlation (the lower row) as functions of $\rho$ for different combinations of angular momentum mode at $\mu=120$ MeV, $240$ MeV, and $360$ MeV, respectively.}
\label{figure7}
\end{figure*}

Figure \ref{figure7} displays the non-Gaussian fluctuations as functions of the radial coordinate $\rho$ for several combinations of angular momentum modes. In all cases, a clear enhancement is observed within the phase transition region. Notably, the correlation amplitudes for nonzero angular momentum modes reach magnitudes comparable to those of the zero mode, consistent with the behavior seen in the two-point correlation functions shown in Figure \ref{figure5}.
The similarity among the two-, three-, and four-point correlations can be attributed primarily to the energy cutoff imposed in the present analysis, which restricts the contributions to the lowest radial mode ($n=1$). As illustrated in subplot (c) of Figure \ref{figure2}, the overall shapes of the probability density are insensitive to the choice of angular momentum modes for the $n=1$ modes.

Based on the above analysis of Gaussian and non-Gaussian anisotropic correlations, the prominence of nonzero $l$ modes signifies that critical fluctuations in a inhomogeneous thermal background carry a distinct spatial anisotropy, encoded in the azimuthal structure of the order parameter field.
This behavior stands in sharp contrast to two well-known limiting cases: (i) a homogeneous system near criticality, where fluctuations are dominated by the zero mode, while contributions from nonzero modes are negligible; and (ii) pure statistical (Gaussian) fluctuations, in which all angular modes are equally weighted and fluctuations are strictly Gaussian, the higher-order fluctuations vanish in this case.

In the present temperature-inhomogeneous setup, the effective potential $V_{\text{eff}}$  suppresses the would-be divergent correlation function via a finite spectral gap and redistributes fluctuation strength among several finite-$l$ channels.
As a result, the fluctuation spectrum retains a measurable anisotropic imprint even after ensemble averaging, which is a direct consequence of the symmetry breaking imposed by the spatially varying background. Such structured fluctuations can, in principle, be transmitted to final-state particle distributions as shown in the next subsection.

\subsection{Relation to the particle number fluctuations}
In this part, we make connections between theoretical predictions and experimental observations.
In relativistic heavy-ion collisions, when the expanding system locally cools to the freeze-out temperature \(T_f\) below the phase transition temperature, particles in the corresponding region decouple from the fireball and propagate to the detector. The spatial fluctuations are mapped onto the momentum spectrum of the final-state particles via the Cooper-Frye formula \cite{Cooper:1974}.

The fluctuations of final-state protons serve as a primary observable for the QCD phase transition signal, because their mass is affected by fluctuations of the \(\sigma\) field through modifications of the effective mass or coupling strength \cite{Stephanov:2008qz}. Assuming a Boltzmann distribution on the freeze-out surface in the local rest frame, the final-state momentum spectrum is related to the local velocity field, the freeze-out temperature, and the freeze-out hypersurface. In our setup, we work in the longitudinal comoving frame and neglect the transverse velocity. In the quasi-steady framework, we simply assume that particles in the freeze-out region are those that reach the detector. The particle number density at the freeze-out region is then
\begin{equation}\label{dis}
n\left( \bm{r}\right) =\int d^3\bm{p}\exp \left\{ -\left[\sqrt{\bm{%
p}^{2}+ g^2(\sigma_0+\sigma)^{2}}-\mu \right]/T_f(\bm{r})\right\}.
\end{equation}%
where the proton mass arises from its coupling to the $\sigma$ field via $g (\sigma+\sigma_0)\bar{p}p$, with $g$ being the coupling strength. $\sigma_0$ (constant) is the shift of the order parameter used in the Ising parameterization, so that the phase transition occurs at $\sigma=0$ \cite{Jiang:2023nmd}.
Expanding the $\sigma$ field as $\sigma = \sigma_c +\tilde\sigma$, the fluctuations in $\sigma $ translate into proton number fluctuations.  To linear order, this yields \cite{Stephanov:2008qz}
\begin{equation}
n\left( \bm{r}\right) = n_c \left( \rho\right) + \delta n (\bm{r})= n_c \left( \rho\right) + \tilde h(\rho)\tilde{\sigma}(\bm{r}),
\end{equation}%
where $n_c$ is obtained by substituting the steady solution $\sigma=\sigma_c$ into equation \eqref{dis} and
\begin{eqnarray}
\tilde h(\rho) &=&-\int d^3 \bm{p} \frac{g^2 (\sigma_0+\sigma_c)}{T_f\left( \rho \right)  \sqrt{\bm{p}^{2}+g^{2}
(\sigma_0+\sigma _{c})^2 }}\times \notag \\
&& \exp \left[ -\frac{\sqrt{\bm{p}^{2}+g^{2}
(\sigma_0+\sigma_{c})^2 }-\mu}{T_f\left( \rho_f \right)} \right].
\end{eqnarray}%

Similar to the above definition of angular momentum mode of $\sigma$ in equation \eqref{sigmal}, we define the fluctuations of proton number at a narrow freeze-out region ($\Delta\rho$) in a specific angular momentum channel to be
\begin{equation}
\delta N_{l}  \equiv \rho \Delta \rho \int dz \int_{0}^{2\pi }d\theta e^{-il\theta }\delta n\left( \bm{r}\right) = h(\rho) \tilde\sigma_l {\big |}_{\rho =\rho_f},
\end{equation}
where $ h(\rho_f) \equiv \tilde h(\rho_f) \rho_f \Delta \rho$    is defined on the freeze-out region and thus takes fixed numerical values for a given temperature profile and freeze-out temperature. Such critical fluctuations induce angular momentum components in the particle number fluctuations, so that \(\delta N_l\) will directly contribute to the experimentally measured anisotropic flow coefficient  \(\nu_l\) .

In non-central collisions, \(\delta N_l\) is dominated by the initial geometric flow. In central collisions, the geometric contribution to \(\langle \delta N_l \rangle\) is averaged out, and the critical fluctuations also do not contribute to \(\langle \delta N_l \rangle\), since \(\tilde\sigma_l(\rho_f) = 0\) for all \(l\ge 1\).  As the critical signals manifest in the variance and non-Gaussian correlations of the flow coefficients, higher-order mode correlations need to be considered.
The two-, three- and four-point correlations of the fluctuating particle number at different $l$-channel are formally expressed as
\begin{eqnarray}
&& \left\langle \delta N_{l_{1}} \delta N_{l_{2}} \right\rangle \propto \left\langle \tilde{\sigma}%
_{l_{1}} \tilde{\sigma}_{l_{2}}\right\rangle, \\
&& \left\langle \delta N_{l_{1}} \delta N_{l_{2}}
\delta N_{l_{3}} \right\rangle  \propto  \Gamma_{l_1l_2l_3}^3(\rho_f),  \\
&& \left\langle \delta N_{l_{1}} \delta N_{l_{2}}
\delta N_{l_{3}} \delta N_{l_{4}}
\right\rangle  \propto \Gamma_{l_1l_2l_3l_4}^4(\rho_f).
\end{eqnarray}%
From the above relations, we see that central collisions offer a particularly clean probe of critical fluctuations, as the main signals arise from the fluctuations themselves, with little background from the mean flow associated with the reaction plane. The higher-order \(\delta N_l\) correlations, such as the three-point and four-point ones that generate non-Gaussianity, are most pronounced in central collisions. We therefore suggest that experimental searches may concentrate on mixed-harmonic correlations in central collisions, such as \(v_l\{4\}\), \(\langle \nu_{l_1} \nu_{l_2} \nu_{l_3} \rangle\) with \(l_1+l_2+l_3=0\), and \(\langle \nu_{l_1} \nu_{l_2} \nu_{l_3} \nu_{l_4} \rangle\) with \(l_1+l_2+l_3+l_4=0\). Meanwhile, correlations that do not satisfy the above angular momentum relations should not yield any flow signal, which can serve as a further cross-check and confirmation of the phase transition signature.

In summary, these anisotropic correlation (with finite $l$) of the $\sigma$ field should in principle imprint themselves in the final-state proton distributions. Beyond the observation of event-by-event multiplicity fluctuations, the phase transition signals can be naturally reflected in measurements of anisotropic flow. Especially, azimuthally sensitive observables, in particular higher-order flow coefficients and their correlations, provide a direct and complementary experimental avenue for detecting the imprint of the QCD phase transition, leveraging precisely the anisotropic enhancement revealed here.

{In practice, extracting critical signals from heavy-ion data requires disentangling them from numerous non-critical effects, such as initial fluctuations \cite{Pang:2012he,Shen:2017bsr}, volume fluctuations \cite{Skokov:2012ds}, dynamical evolution \cite{Stephanov:2017ghc,Du:2020bxp,An:2020vri}, and the freeze-out procedure \cite{Pradeep:2022eil,Pradeep:2022mkf}. This can be achieved by comparing the experimental data with the predictions of a well-modeled non-critical background. Our analysis yields explicit predictions for the critical correlations of the \(\sigma\)  field. Their characteristic angular-momentum signatures, including specific selection rules and non-Gaussian cumulants, provide a clear means to isolate them from non-critical contributions.}

\section{Summary and outlook}  \label{sec5}

In this work, we have developed a quantum statistical framework for investigating the locally equilibrated  fluctuations and spatial correlations of the chiral order parameter field within the Ising model for a system featuring a spatially non-uniform temperature profile, as motivated by the conditions in relativistic heavy-ion collisions. By constructing the partition function and the corresponding effective action for the $\sigma$ field in a spatially non-uniform temperature background, we derived the base field configuration and analyzed the two-point fluctuations perturbatively. We further extract the single angular momentum mode and connect it with the experimental observables.

A central outcome of this formalism is the explicit identification of the system's possible excitation modes. For a given energy cutoff, the spectrum of low-energy fluctuations can be systematically decomposed into a set of eigen-modes characterized by radial ($n$) and angular momentum ($l$) quantum numbers. This decomposition allows for a direct, mode-by-mode analysis of fluctuation contributions, revealing that the critical region primarily excites modes localized near the phase-transition interface.

Our results demonstrate that the spatial pattern of fluctuations is fundamentally shaped by the thermal geometry of the system. Specifically, nonlocal correlations exhibit marked anisotropy: for a single mode with specific $n$ and $l$, they maintain long-range coherence along isothermal contours, while fluctuations in the radial direction are strongly suppressed due to the temperature inhomogeneity. Importantly, the cumulative summation of these distinct eigen-modes progressively breaks the initial long-range azimuthal coherence, causing the correlation strength to become increasingly localized around a fixed anchor point. This transition from long-range coherence to localized correlation patterns illustrates how mode superposition reshapes the overall correlation structure of the system.

Furthermore, we have extracted and analyzed the individual contributions of non-zero angular momentum modes, which exhibit mutually comparable correlation strengths. From a physical perspective, the spatially non-uniform temperature distribution introduces a continuous energy scale, and naturally alters the distribution of fluctuation modes, transforming the system from a regime dominated by the zero momentum mode in uniform conditions to a heterogeneous regime where multiple angular momentum channels participate with comparable weight. By constructing observables analogous to experimental anisotropic flow coefficients, we have directly linked these structured, finite-$l$ fluctuations to measurable quantities in heavy-ion collisions, thereby establishing a concrete bridge between our theoretical framework and potential experimental signatures.

In conclusion, the formalism presented here offers a new perspective on how thermal geometry reshapes fluctuation patterns in inhomogeneous systems near phase transitions.
Our findings suggest that azimuthally sensitive observables, particularly anisotropic flow fluctuations, may provide a unique and complementary probe of critical phenomena.
While the current framework assumes rotational symmetry, we will subsequently extend it to systems with elliptic symmetry to disentangle the anisotropy of critical fluctuations from the initial-state anisotropy.
Looking ahead, to enable quantitative comparisons with experimental data from the Beam Energy Scan program, future work will incorporate real-time dynamical evolution and a more explicit coupling to the hydrodynamic background.

\section*{Acknowledgments}
We thank Xiaofeng Luo, Nu Xu, Yi Yin, Huichao Song, Yu-xin Liu for uesful discussions and comments.
L. Jiang acknowledges the support from the National Natural Science Foundation of China (NSFC) under grant No.\,12105223 and No.\,12247103, and Shaanxi Fundamental Science Research Project for Mathematics and Physics under grant No.\,25JSQ031, T. Yang is supported by the National  Science Foundation of China under Grants  No.\,12575029 and  No.\,12175180,
and J.-H. Zheng acknowledges the support from the NSFC under Grant No.\,12575028 and Natural Science Basic Research Program of Shaanxi Province under Grant No.\,2024JC-YBMS-022.



%

\end{document}